# Terahertz magneto-optical spectroscopy of two-dimensional hole and electron systems

N. Kamaraju[1], W. Pan[2], U. Ekenberg[3], D. M. Gvozdić[4], S. Boubanga-Tombet[1,5], P. C. Upadhya[1,6], J. Reno[2], A. J. Taylor[1], and R. P. Prasankumar[1]

[1] Center for Integrated Nanotechnologies, Los Alamos National Laboratory, Los Alamos, NM 87545, USA

[2] Sandia National Laboratories, Albuquerque, NM 87123, USA

[3] Semiconsultants, Brunnsgränd 12, SE-18773 Täby, Sweden

[4] School of Electrical Engineering, University of Belgrade, Belgrade 11120, Serbia

[5] Research Institute of Electrical Communication, Tohoku University, 2-1-1 Katahira, Aoba-Ku, Sendai, Japan

[6] Laboratory for Electro-Optics Systems, Indian Space Research Organization, Bangalore - 560058, India

We have used terahertz (THz) magneto-optical spectroscopy to investigate the cyclotron resonance in high mobility two-dimensional electron and hole systems. Our experiments reveal long-lived (~20 ps) coherent oscillations in the measured signal in the presence of a perpendicular magnetic field. The cyclotron frequency extracted from the oscillations varies linearly with magnetic field for a two-dimensional electron gas (2DEG), as expected. However, we find that the complex non-parabolic valence band structure in a two-dimensional hole gas (2DHG) causes the cyclotron frequency and effective mass to vary nonlinearly with the magnetic field, as verified by multiband Landau level calculations. This is the first time that THz magneto-optical spectroscopy has been used to study 2DHG, and we expect that these results will motivate further studies of these unique 2D nanosystems.

PACS number(s): 78.67.De, 73.20.−r, 76.40.+b, 73.43.-f, 78.67.-n

Two-dimensional electron and hole gases (2DEGs and 2DHGs) in high mobility semiconductor quantum wells are well known to be an ideal playground for exploring fascinating quantum phenomena. These include the quantum Hall effect [1] (QHE), 2D metal-insulator transitions [2], composite fermions [3], and high magnetic field-induced Wigner crystal phases [4], making them the focus of intense study for several decades. Furthermore, they can be used as high electron mobility transistors and to detect infrared (IR) [5] and THz [6] radiation, as well as in fundamental studies of cavity quantum electrodynamics [7].

Previous studies have mainly been focused on 2DEGs, where the highest electron mobility is achieved. However, in recent years, it has become clear that their use in the burgeoning areas of quantum computing and spintronics will be limited due to the strong interactions between electron spins (in an *s*-type conduction band) and nuclei. This makes spin coherence very fragile, which can destroy information before it can be processed [8,9]. In this regard, a 2DHG would have an advantage, as the *p*-type hole valence band has a reduced hyperfine interaction with the nuclei, leading to longer spin coherence times for holes [10,11]. Furthermore, the Rashba effect (electric-field-induced spin splitting) can be made much stronger in a 2DHG than in a 2DEG [12]. Consequently, 2DHGs are more interesting for future spintronics and quantum information applications [13], making it particularly important to explore their physics in depth. Another significant difference between 2DHGs and 2DEGs is that the hole effective mass, $m_{eff}^h$, is ~3-12 times the electron effective mass ($m_{eff}^e$), which causes a corresponding reduction in the cyclotron resonance frequency ($\omega_c = eB/m_{eff}^h$, where *e* is the electron charge and *B* is the magnetic field), Landau level splitting (given by $\hbar\omega_c$) and carrier mobility ($\mu=\tau e/m_{eff}^h$), where $\tau$ is the carrier scattering time.

In addition to electronic transport studies of 2DEGs/2DHGs, infrared and microwave radiation [14-18] have been utilized to study some of the quantum phenomena described above, where the quantum coherence is generated and investigated in the presence of a perpendicular magnetic field. However, the cyclotron frequency $\omega_c$ in 2DHG is less than 1 THz throughout a large magnetic field range, making it difficult to measure using conventional spectroscopic techniques [15]. Terahertz time-domain spectroscopy (THz-TDS) is an attractive alternative for studying the properties of low-dimensional quantum nanostructures, as THz radiation (frequency range ~0.1-10 THz) has been shown to directly probe low energy excitations, such as phonons, excitons, and Cooper pairs [19]. More recently, THz-TDS has also been used to study the cyclotron frequency and the QHE in 2DEGs (composed of e.g., doped GaAs quantum wells [20-25] and graphene [26,27]), since the THz photon energy is on the order of $\hbar\omega_c$ in these systems. However, to the best of our knowledge, there have been no THz studies on 2DHG reported to date. This could be a fruitful direction, since the complex valence band structure in semiconductor quantum wells will likely cause quantum phenomena such as the cyclotron resonance and QHE to exhibit some unique features in 2DHG as compared to 2DEG.

Here, we used THz-TDS to measure the cyclotron resonance frequency, effective mass, and carrier scattering time in a high mobility two-dimensional hole system under a quantizing magnetic field. Unlike IR or microwave spectroscopy, THz-TDS yields both the amplitude and phase of the THz electric field transmitted through the sample [28], allowing us to directly measure the cyclotron frequency and the scattering time in the time domain. Furthermore, we concurrently performed these experiments on both 2DEG and 2DHG to give more insight into their physics.

The samples used in our experiments were grown using molecular beam epitaxy on a 625 μm thick (001) GaAs wafer. The two-dimensional electron (hole) gas layer consists of a 30 (15) nm thick GaAs quantum well, modulation doped (using silicon for the 2DEG and carbon for the 2DHG) to a carrier density, $n_{2D}$, of ~1.8 x $10^{11}$ cm$^{-2}$ (2DEG) and ~1.4 x $10^{11}$ cm$^{-2}$ (2DHG) at 1.5 K and 0.3 K, respectively. The mobilities ($\mu$) of the 2DEG and the 2DHG are 3 x $10^6$ cm$^2$/Vs and 0.3 x $10^6$ cm$^2$/Vs, respectively. Fig. 1 displays the results of these DC measurements for both samples. Well-developed quantum Hall states are observed in both samples. From the plateau structures seen in these measurements, we could identify the corresponding Landau level filling factor, $\nu = \frac{h n_{2d}}{e * B}$, as marked in Figures 1(a)-(b). For the THz measurements, two samples (2DEG and 2DHG) were cut from the same wafers. Both samples had a large area of 10 mm x 10 mm, so that the focused THz spot (~4 mm) probes the center of the sample and avoids artifacts due to diffraction.

Coherent single-cycle THz pulses were generated by focusing 50 femtosecond (fs) pulses at 800 nm from an 80 MHz Ti:Sapphire oscillator on a +30 V DC biased low temperature-grown GaAs photoconductive dipole antenna. The resulting THz radiation is centered at 0.8 THz, with a bandwidth of 1.3 THz. The sample was then placed in a split-coil superconducting magnet, allowing us to vary the magnetic field perpendicular to the sample surface (Faraday geometry) from 0 T to ± 6 T. This THz magneto-optical spectroscopy setup allows us to perform polarization-dependent THz-TDS experiments under a magnetic field [20-25] and measure the Faraday rotation of the transmitted THz pulse. All measurements were performed at 1.8 K ($k_B T$~0.03 THz). The typical time-domain THz electric field with no sample and its corresponding frequency spectrum, along with a more detailed description of the experimental technique and setup, is given in the Supplementary Material.

Our first THz magneto-optical measurements were performed on the 2DEG sample, for comparison to previous studies [20-25] With the applied field, *B*, perpendicular to the sample (in the *z* direction) and the incident THz polarization in the *x*-direction (see the supplementary material), the electrons undergo cyclotron motion around *B*, which is reflected in the Faraday rotation induced in the polarization of the transmitted THz pulse. This cyclotron motion (Faraday rotation) is an odd function of *B*, meaning that the direction of motion reverses when *B* is reversed. Thus, the induced THz transients along the *y* axis due to the cyclotron resonance in the 2DEG [22] can be obtained by subtracting the transmitted THz waveform at –*B* from the waveform at +*B* [24] and dividing the result by two. This eliminates any spurious response of the photoconductive antenna to the magnetic field, as well as any artifacts due to the low THz polarization extinction ratio (100:1) of this detection method [24], ensuring that the subtracted signal mainly contains oscillations induced through the Faraday effect. The resultant subtracted THz electric fields, $E_y^c(t) \equiv ([E_y(B,t) - E_y(-B,t)]/2)$ for selected DC magnetic fields are shown in Figure 2(a). Clear oscillations are observed that last much longer than the incident THz pulse duration (Fig. 2(a)), which are linked to the high mobility of the 2DEG, as will be shown below. The measured cyclotron oscillations can be fit using the equation

$$E_y^c(t) = E_0 + A_0 * exp\left(-\frac{t}{\tau_{CR}}\right) * sin(\omega_c t), \qquad (1)$$

where $\tau_{CR}$ is the scattering time, $E_0$ is the DC offset and $A_0$ is the amplitude of the oscillations. The fit agrees well with the data, and the fit parameters $\omega_c$ and $\tau_{CR}$, are plotted as a function of $B$ in Figs. 2(b) and (d). It can be seen that the cyclotron frequency varies linearly with magnetic field, as expected from the parabolic band structure of the 2DEG. A linear fit to the data in Fig. 2(b) allows us to obtain $m_{eff}^e \cong 0.072 m_e$, where $m_e$ is the bare electron mass (Fig 2(c)).

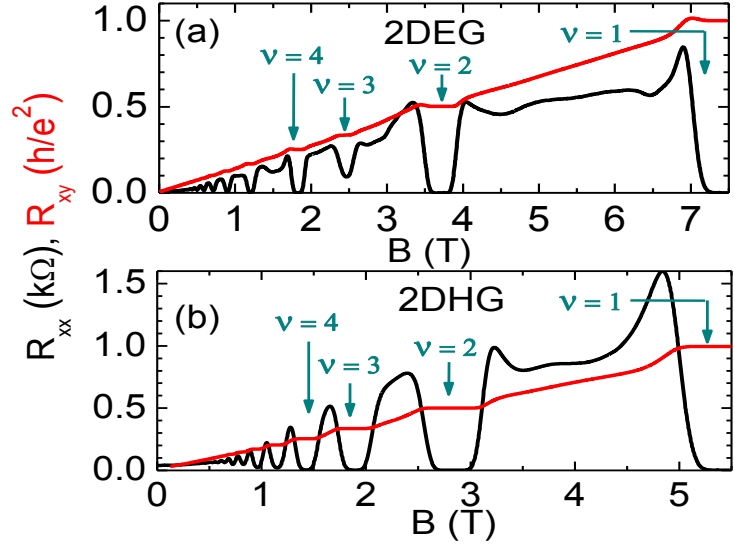

Figure 1: (a) DC Hall resistivity, $R_{xx}$ and $R_{xy}$, as a function of applied magnetic field, B, for the 2DEG, measured at 1.5 K. The Landau level filling factors of $\nu$ = 1, 2, 3, and 4 are indicated on the figure. (b) Similar DC measurements for the 2DHG measured at 0.3 K, with Landau level filling factors also marked.

One advantage of THz-TDS over other experimental techniques is that it allows us to directly measure $\tau_{CR}$ in the time domain. Figure 2(d) indicates that $\tau_{CR}$ in the 2DEG is influenced by the magnetic field, varying between ~28 ps at $B$=0.25 T to ~8 ps at $B$=4 T. Moreover, visible oscillations in $\tau_{CR}$ are seen in Figure 2(d), as predicted theoretically (i.e., the e-e scattering time oscillates with the LL filling factor [29,30]). These observations are different from those reported in ref. [25], where $\tau_{CR}$ displayed a weak and smooth dependence on $B$. We believe that this difference is probably due to the lower measurement temperature in our experiment and consequently better-developed quantum Hall effect. These oscillations in $\tau_{CR}$ are believed to occur due to the screening of impurities, which is maximized when a LL is half filled (increasing $\tau_{CR}$) and minimized when it is filled [29,30] (decreasing $\tau_{CR}$). A precise identification of a local minimum in $\tau_{CR}$ with respect to a specific Landau level filling is not possible at the present time, since the DC transport measurements were carried out in a different sample (though cut from the same wafer) and in a different cryogenic system. Nevertheless, the local minima at $B$ = 2.5, 1.5, and 0.75 T correspond to Landau level filling factors of $\nu$ = 3, 5, and 10, respectively, if we assume the two samples have the same electron density. Finally, we can also estimate the mobility, $\mu$, using the values obtained for $m_{eff}^e$ and $\tau_{CR}$. This gives $\mu$ ~ 7.3 x $10^5$ cm$^2$/Vs at low magnetic fields, which is approximately a quarter of the mobility from DC measurements, 3 x $10^6$ cm$^2$/Vs at 0 T and 1.5 K. A similar discrepancy was also reported recently [25]. There, it was believed that the cyclotron mobility is dominated by cooperative radiative decay, which is much faster than static impurity scattering processes; this could also be occurring here, although further investigation is required to confirm this.

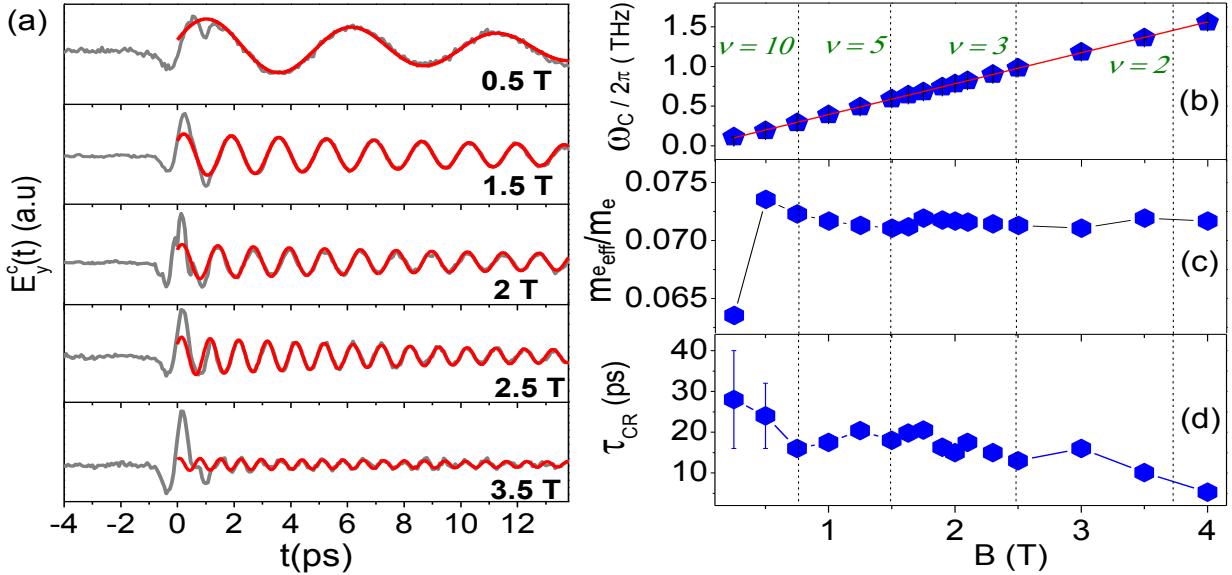

Figure 2: (a) The difference in the THz electric fields transmitted through the 2DEG for $\pm B$, detected in the $y$ direction by keeping the second WGP at 135°. (b, c, d) Magnetic field dependence of $\omega_c/2\pi$, $m_{eff}^e$, and $\tau_{CR}$. The red lines show fits to the data, as described in the text, and the dashed vertical lines represent filling factors, $\nu = 2, 3, 5, 10$.

We then performed similar measurements on the 2DHG; the THz transients induced in the $y$-direction, $E_y^c$, are plotted vs. magnetic field in Figure 3(a). A good fit to the data was obtained using equation (1), as for the 2DEG. The corresponding fit parameters are plotted in Fig. 3(b-d). As expected, the cyclotron frequencies in the 2DHG are almost four times smaller than in the 2DEG, due to the higher hole effective mass in this sample. The most noteworthy feature in this data is that $\omega_c$ varies linearly with $B$ at low fields (<4 T), but as the magnetic field increases further, $\omega_c$ begins to saturate. As we will show below, this is due to the non-parabolic band structure of the 2DHG, which is influenced by the interactions between light hole (LH) and heavy hole (HH) valence bands, as compared to the parabolic subband structure in the conduction band of the 2DEG.

In addition, the magnetic field dependence of $\tau_{CR}$ in the 2DHG is somewhat different from what was observed in the 2DEG. It drops quickly from ~ 40 ps at $B$ ~ 1 T to ~8 ps at 2 T, increases from 2 to 3.5 T, remains roughly constant from 3.5 to 5.5 T, and increases again at higher magnetic fields. This irregular dependence may be related to the contributions of different Landau level transitions to the measured cyclotron frequency in the 2DHG, as discussed below. Two points related to Figure 3(d) are worth emphasizing. First, in the low field regime ($B < 2$ T), $\tau_{CR}$ for the 2DHG is relatively large, even larger than that in the 2DEG. Second, $\mu_{CR}$ in the 2DHG is closer to $\mu_{DC}$. From the initial linear fit to $\omega_c$ vs. $B$ at lower fields (up to 4 T), $m_{eff}^h \cong$ 0.27$m_e$ is obtained. This, combined with $\tau_{CR}$, yields a $\mu_{CR}$ of ~ 0.3 x $10^6$ cm$^2$/Vs at $B = 1$

T, which agrees with the measured DC mobility of $0.3 \times 10^6$ cm$^2$/Vs at 0 T and 1 K. This is dramatically different from the case in the 2DEG, where $\mu_{CR} \ll \mu_{DC}$. At the present time, the reason for this discrepancy is not known. One possibility is that in the 2DHG the transverse relaxation process is dominant, rendering the CR and DC mobility the same [25].

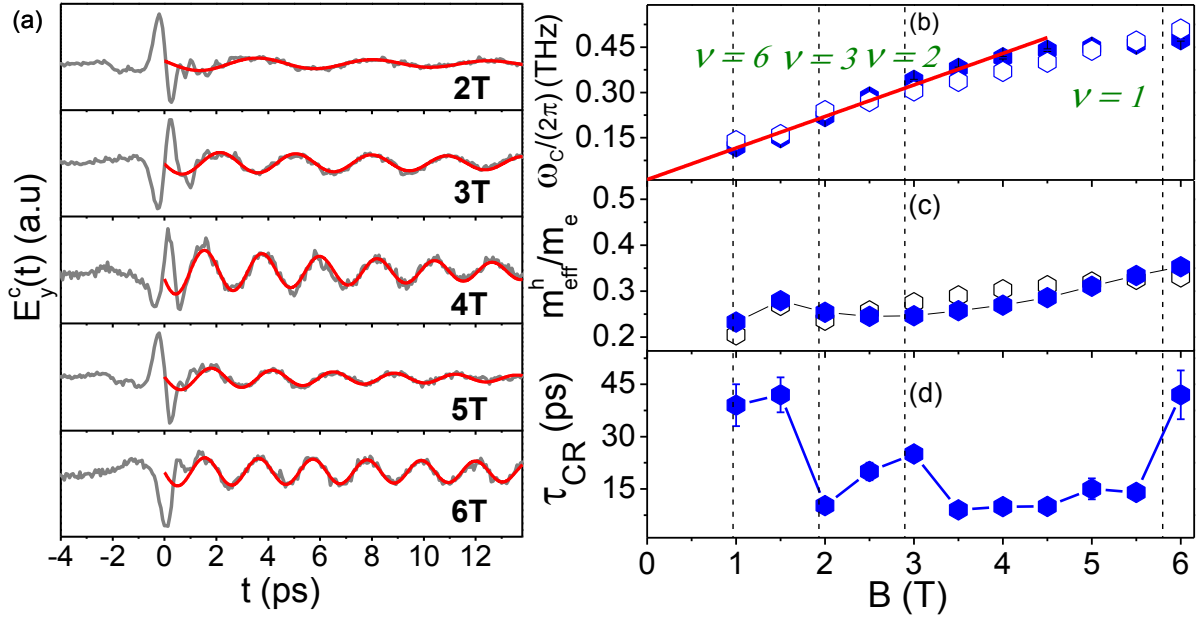

Figure 3: (a) The difference in the THz electric fields transmitted through the 2DHG for $\pm B$, detected in the $y$ direction. (b, c, d) Magnetic field dependence of $\omega_c$, $m_{eff}^h$ and $\tau_{CR}$. In (a), the red lines show a sinusoidal fit to the data, as described in the text. In (b) the red line shows a linear fit for lower $B$ values, up to 4 T. The open symbols in (b) are theoretically calculated, as described in the text, and the dashed vertical lines represent filling factors, $\nu = 1, 2, 3, 6$.

To obtain a deeper understanding of the nonlinear variation of the cyclotron frequency at high magnetic fields in the 2DHG, we have performed self-consistent Hartree calculations of the electronic subband structure, which is then used to calculate the Landau levels as a function of magnetic field, as in ref. [31]. We can then determine the energies of the allowed transitions, allowing us to calculate the cyclotron frequency and effective mass for a given value of $B$ for comparison to the experimental results. We used a multiband envelope function approach to calculate the subband structure in the 2DHG. From the slopes of the subbands at the Fermi level, one can determine the semiclassical cyclotron masses. Assuming non-parabolic subbands, we then obtained an approximate effective mass of 0.28m$_e$ from the slope of the highest HH band by averaging over the [11] and [10] directions. However, as was shown in ref. [31], the semiclassical masses correspond to the limit $B \rightarrow 0$ and thus cannot explain the variation of the cyclotron mass with $B$.

For more insight into this, we performed Landau level calculations, using a simplified approach that was based on a 4x4 Hamiltonian matrix including the heavy hole and light hole bands, along with spin, in the axial approximation. The split-off and conduction bands were included perturbatively. The Landau level structure for the 2DHG is shown in Figure 4. It is clearly seen that most of the Landau levels vary nonlinearly with magnetic field. Finally, the cyclotron transition energies were calculated by applying the standard selection rules [30,31]. A complication in the 2DHG system is that as the filling factor varies, different transitions, corresponding to different masses, become possible. For holes, it is common to label the Landau level indices starting from -2. The calculated frequencies are shown as open symbols in Fig. 3(b), revealing good agreement with the experimentally measured cyclotron frequencies. This also enables us to determine that the cyclotron resonance is due to the transition from LL1 to LL2 for fields above 2 T. At 1.5 T, the 2-3 transition is expected to take place, and at 1 T, both the 0-1 and 3-4 transitions contribute to the measured cyclotron frequency. Overall, the agreement between the measured and calculated cyclotron frequencies supports the idea that the non-parabolicity of the heavy and light hole subbands causes the measured cyclotron frequency and effective mass to depend nonlinearly on $B$ (Figs. 3(b) and (c)), in agreement with the results of ref. 16 and in contrast with our results on the 2DEG.

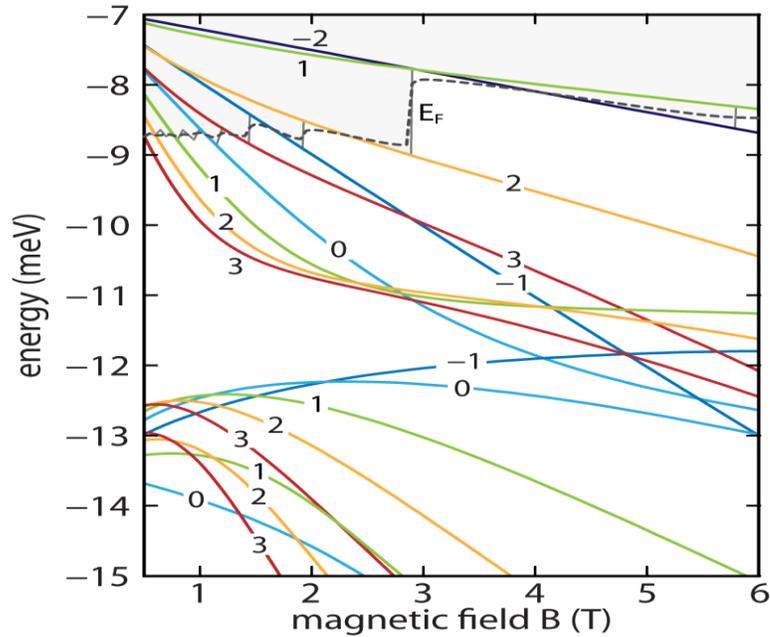

Figure 4: Calculated variation of the Landau level energies with magnetic field. The variation of the Fermi level, $E_F$, with magnetic field is indicated by the dashed lines (including broadening) and solid vertical lines (without broadening). The upper levels (approximately from -10 to -7 meV for $B < 1$ T) originate from the heavy hole band, with the lower levels (approximately from -14 meV to -12 meV for $B < 1$ T) originating from the light hole band.

In conclusion, we used THz magneto-optical spectroscopy to reveal the nonlinear dependence of the cyclotron resonance frequency and effective mass on magnetic field in a 2DHG, unlike the linear dependence seen in a 2DEG. This is likely due to the existence of coupled non-parabolic valence bands in the 2DHG, causing the transition energies between them to vary nonlinearly with magnetic field, as verified by our calculations. We also observed weak oscillations in the carrier scattering time with increasing magnetic field for both the 2DEG and 2DHG. To the best of our knowledge, this is the first measurement of these parameters as a function of magnetic field in a 2DHG using THz-TDS, shedding light on the novel properties of 2DHGs that should stimulate further studies of these unique nanosystems.

# Terahertz magneto-optical spectroscopy of two-dimensional hole and electron systems

N. Kamaraju[1], W. Pan[2], U. Ekenberg[3], D. M. Gvozdić[4], S. Boubanga-Tombet[1,5], P. C. Upadhya[1,6], J. Reno[2], A. J. Taylor[1], and R. P. Prasankumar[1]

[1] Center for Integrated Nanotechnologies, Los Alamos National Laboratory, Los Alamos, NM 87545, USA

[2] Sandia National Laboratories, Albuquerque, NM 87123, USA

[3] Semiconsultants, Brunnsgränd 12, SE-18773 Täby, Sweden

[4] School of Electrical Engineering, University of Belgrade, Belgrade 11120, Serbia

[5] Research Institute of Electrical Communication, Tohoku University, 2-1-1 Katahira, Aoba-Ku, Sendai, Japan

[6] Laboratory for Electro-Optics Systems, Indian Space Research Organization, Bangalore - 560058, India

**Experimental details:**

In our magneto-optical THz-TDS experiments, coherent single-cycle THz pulses are generated by focusing 50 femtosecond (fs) pulses at 800 nm from an 80 MHz Ti:Sapphire oscillator on a +30 V DC biased low temperature-grown GaAs photoconductive dipole antenna, using a short focal length lens with an average power of 10 mW. The experimental setup is shown in Fig. S1. THz radiation peaked at 0.8 THz, with a FWHM of 1.3 THz, is generated from PCA1 and focused onto the sample by a pair of parabolic mirrors (OAP1 and OAP2). The THz transmission after the sample was collimated and re-focused onto a similar photoconductive antenna (PCA2) using another pair of parabolic mirrors (OAP3 and OAP4), allowing us to sample the THz electric field.

The sample was placed in a split-coil superconducting magnet, allowing us to vary the magnetic field perpendicular to the sample surface (Faraday geometry) from 0 T to ± 6 T. In this geometry, one wire grid polarizer (WGP) is kept before the sample to set the input polarization at 45° ('*x*' axis) with respect to the optical table and a second wire-grid polarizer is kept at either 45° or 135° ('*y*' axis) to record both parallel and perpendicular components of the transmitted THz field. In the presence of a magnetic field, a THz electric field incident along the *x* axis (at 45°) induces an orthogonal component of the transmitted electric field along the *y* axis (at 135°). The typical time-domain THz electric field with no sample and its corresponding frequency spectrum is shown in Figure S2.

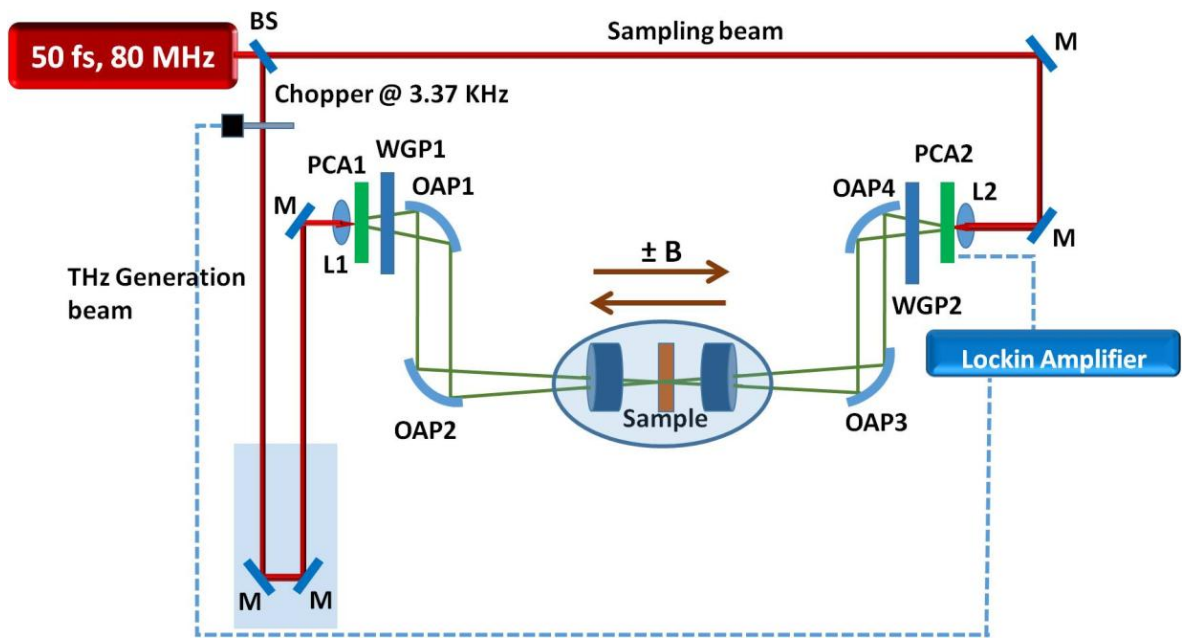

Fig. S1. (color online) Schematic of the experimental setup to measure THz Faraday rotation. M: Mirrors, BS: Beam splitter, L1-2: Short focal length lenses, PCA1-2: Photoconductive dipole antennas, OAP1-4: Off-axis parabolic mirrors, WGP1-2: Wire grid polarizers.

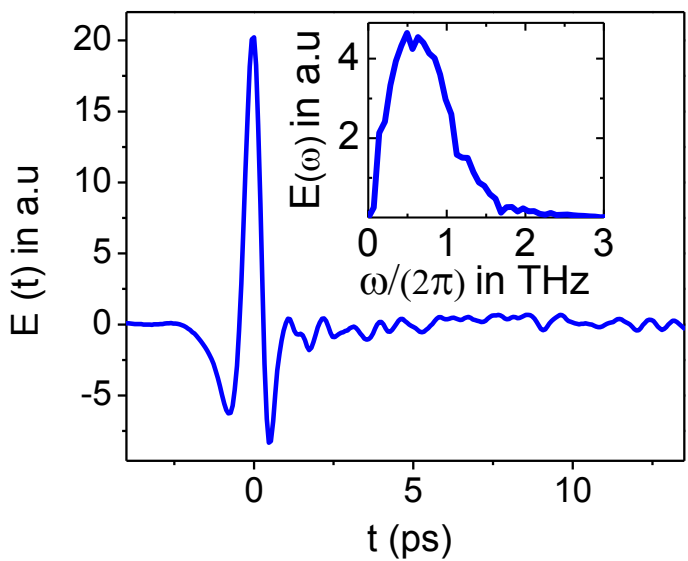

Fig. S2. (color online) Terahertz electric field with no sample. The inset shows the frequency spectrum, obtained by Fourier transforming the time-dependent electric field.